%% file: main.tex
 \documentclass[sigconf]{acmart}
\AtBeginDocument{%
  }
    
\usepackage[turnon]{notes}
\input{includes/includes}

\input{includes/pygments}
\input{includes/macros}

\newcommand{\cz}[1]{{\color{orange}[Chengzhi says: #1]}}
\SetKwInOut{KwIn}{Input}
\SetKwInOut{KwOut}{Output}
\setcopyright{acmcopyright}

\copyrightyear{2024}
\acmYear{2024}
\setcopyright{rightsretained}
\acmConference[ICSE '24]{2024 IEEE/ACM 46th International Conference on Software Engineering}{April 14--20, 2024}{Lisbon, Portugal}
\acmBooktitle{2024 IEEE/ACM 46th International Conference on Software Engineering (ICSE '24), April 14--20, 2024, Lisbon, Portugal}\acmDOI{10.1145/3597503.3639170}
\acmISBN{979-8-4007-0217-4/24/04}

\begin{document}

\title{Towards Causal Deep Learning for Vulnerability Detection}

\author{Md Mahbubur Rahman}
\email{mdrahman@iastate.edu}
\orcid{0009-0000-2680-2309}

\affiliation{%
  \institution{Iowa State University}
  \streetaddress{Atanasoff Hall, 2434 Osborn Dr}
  \city{Ames}
  \state{IA}
  \country{USA}
  \postcode{50011}
}

\author{Ira Ceka}
\email{ira.ceka@columbia.edu}
\affiliation{%
  \institution{Columbia University}
  \city{New York}
  \state{NY}
  \country{USA}
}

\author{Chengzhi Mao}
\email{mcz@cs.columbia.edu}
\affiliation{%
  \institution{Columbia University}
  \city{New York}
  \state{NY}
  \country{USA}
}

\author{Saikat Chakraborty}
\email{saikatc@microsoft.com}
\affiliation{%
  \institution{Microsoft Research}
  \city{Redmond}
  \state{WA}
  \country{USA}
}

\author{Baishakhi Ray}
\email{rayb@cs.columbia.edu}
\affiliation{%
  \institution{Columbia University}
  \city{New York}
  \state{NY}
  \country{USA}
}

\author{Wei Le}
\email{weile@iastate.edu}
\affiliation{%
  \institution{Iowa State University}
  \streetaddress{Atanasoff Hall, 2434 Osborn Dr}
  \city{Ames}
  \state{IA}
  \country{USA}
  \postcode{50011}
}


\input{body/0.abstract.tex}


\newcommand{\tool}{{CausalVul}\xspace}
\newcommand{\mycomment}[1]{}



\maketitle

\input{body/1.introduction.tex}
\input{body/2.overview.tex}
\input{body/3.spuriousfeatures.tex}
\input{body/3.causality.tex}

\input{body/3.alg.tex}

\input{body/4.experimental_setup.tex}

\input{body/5.results.tex}
\input{body/7.threats.tex}
\input{body/6.related.tex}

\input{body/8.conclusion.tex}
\bibliographystyle{ACM-Reference-Format}
\bibliography{bibliography}
\end{document}

%% file: includes/includes.tex
\usepackage{hyperref}
\usepackage{boldline}
\usepackage{color,colortbl}
\usepackage{bigstrut}
\usepackage[linesnumbered, ruled]{algorithm2e}
\usepackage{amsmath,amsfonts}
\usepackage{amsmath}
\usepackage{array}
\usepackage{algorithmic}
\usepackage{balance}
\usepackage{blindtext}
\usepackage{booktabs}
\usepackage{caption}
\usepackage{hyperref}
\usepackage[noabbrev]{cleveref}
\usepackage{comment}
\usepackage{enumitem}
\usepackage{eqparbox}
\usepackage{fancybox}
\usepackage{fancyvrb}
\usepackage{framed}
\usepackage{graphicx}
\usepackage{ifthen}
\usepackage{listings}
\usepackage{mathrsfs}
\usepackage{mdwmath}
\usepackage{mdwtab}
\usepackage{multirow}
\usepackage{pifont}
\usepackage{textcomp}
\usepackage{url}
\usepackage{xspace}
\usepackage[normalem]{ulem}
\usepackage[framemethod=TikZ]{mdframed}
\usepackage{caption}
\usepackage{subcaption}
\usepackage{tabularx}
\usepackage{tcolorbox}
\tcbuselibrary{listings,skins}
\usepackage{mathtools}
\usepackage{blindtext}
\usepackage{lstautogobble}
\DeclareGraphicsExtensions{.pdf,.jpeg,.png}
\graphicspath{{figures/}}

\newlength\Linewidth
\def\findlength{\setlength\Linewidth\linewidth
\addtolength\Linewidth{-4\fboxrule}
\addtolength\Linewidth{-3\fboxsep}
}
\newenvironment{examplebox}{\par\begingroup
  \setlength{\fboxsep}{3pt}\findlength
  \setbox0=\vbox\bgroup\noindent
  \hsize=0.90\linewidth
  \begin{minipage}{0.90\linewidth}\normalsize}
    {\end{minipage}\egroup
    \textcolor{gray20}{\fboxsep1.2pt\fbox
     {\fboxsep5pt\colorbox{gray05}{\normalcolor\box0}}}
    \endgroup\par\noindent
    \normalcolor\ignorespacesafterend}

\newcounter{RQACounter}

\usetikzlibrary{shadows}
\usepackage{graphics}
\newmdenv[
    tikzsetting= {fill=blueish},
    skipabove=0.33em,
    skipbelow=0.33em,
    linewidth=1pt,
    innerleftmargin=4pt,
    innerrightmargin=4pt,
    innertopmargin=2pt,
    innerbottommargin=2pt,
    linecolor=gray95,
    roundcorner=2pt, 
    shadow=true,
    shadowsize=4pt,
    shadowcolor=gray95
]{questionbox}

\newmdenv[
    tikzsetting= {fill=greenish},
    skipabove=0.33em,
    skipbelow=0.33em,
    linewidth=1pt,
    innerleftmargin=4pt,
    innerrightmargin=4pt,
    innertopmargin=2pt,
    innerbottommargin=2pt,
    linecolor=gray95,
    roundcorner=2pt, 
    shadow=true,
    shadowsize=4pt,
    shadowcolor=gray95
]{answerbox}

\newmdenv[
    skipabove=0.33em,
    skipbelow=0.33em,
    innerleftmargin=4pt,
    innerrightmargin=4pt,
    innertopmargin=2pt,
    innerbottommargin=2pt,
]{lessonbox}

\usepackage{tikz}
\newenvironment{lesson}
{
    \begin{lessonbox}
}
{\end{lessonbox}}

\newenvironment{result}
{\begin{answerbox}}
{\end{answerbox}}

\newenvironment{question}
{\begin{questionbox}}
{\end{questionbox}}

\definecolor{javared}{rgb}{0.6,0,0} 
\definecolor{javagreen}{rgb}{0.25,0.5,0.35} 
\definecolor{javapurple}{rgb}{0.5,0,0.35} 
\definecolor{javadocblue}{rgb}{0.25,0.35,0.75} 

\lstdefinestyle{basejava}{
  language=java,
  showstringspaces=false,
  basicstyle=\small\ttfamily,
  keywordstyle=\bfseries\color{javapurple},
  commentstyle=\itshape\blue,
  identifierstyle=\blue,
  frame=none,
  backgroundcolor=\color{white},
}

\lstdefinestyle{CustomJava}{
  belowcaptionskip=\baselineskip,
  breaklines=true,
  xleftmargin=\parindent,
  language=java,
  showstringspaces=false,
  basicstyle=\small\ttfamily,
  keywordstyle=\bfseries\color{javapurple},
  commentstyle=\itshape\blue,
  identifierstyle=\blue,
  belowskip=1pt,
  numbers=left,
  gobble=0
}

\lstdefinestyle{CustomJavaWoNumbers}{
  belowcaptionskip=0.5\baselineskip,
  breaklines=true,
  xleftmargin=\parindent,
  language=java,
  showstringspaces=false,
  basicstyle=\small\ttfamily,
  keywordstyle=\bfseries\color{javapurple},
  commentstyle=\itshape\blue,
  identifierstyle=\blue,
  belowskip=0.5pt,
  numbers=none,
  gobble=0
}

%% file: includes/pygments.tex
\newcommand\blue[1]{\textcolor[rgb]{0.00,0.00,1.00}{{#1}}}

\definecolor{blueish}{RGB}{250, 250, 255}
\definecolor{greenish}{RGB}{250, 255, 250}
\definecolor{redish}{RGB}{255, 200, 200}

\definecolor{highlight}{RGB}{175, 255, 100}
\definecolor{gray01}{gray}{.98}
\definecolor{gray05}{gray}{0.95}
\definecolor{gray08}{gray}{0.92}
\definecolor{gray10}{gray}{0.90}
\definecolor{gray12}{gray}{0.88}
\definecolor{gray15}{gray}{0.85}
\definecolor{gray18}{gray}{0.82}
\definecolor{gray20}{gray}{0.80}
\definecolor{gray25}{gray}{0.75}
\definecolor{gray30}{gray}{0.70}
\definecolor{gray35}{gray}{0.65}
\definecolor{gray40}{gray}{0.60}
\definecolor{gray45}{gray}{0.55}
\definecolor{gray50}{gray}{0.50}
\definecolor{gray55}{gray}{0.45}
\definecolor{gray60}{gray}{0.40}
\definecolor{gray65}{gray}{0.35}
\definecolor{gray70}{gray}{0.30}
\definecolor{gray75}{gray}{0.25}
\definecolor{gray80}{gray}{0.20}
\definecolor{gray85}{gray}{0.15}
\definecolor{gray90}{gray}{0.10}
\definecolor{gray95}{gray}{0.05}

%% file: includes/macros.tex
\xspace%

\newcommand{\Comment}[1]{}

\newcommand{\rasel}[1]{{\todo{Russell:  {\color{orange} #1}}}}
\newcommand{\wei}[1]{{\todo{Wei:  {\color{purple} #1}}}}
\newcommand{\ira}[1]{{\todo{Ira:  {\color{Brown} #1}}}}

\newcommand{\linecode}[1]{\lstinline[escapechar=@,basicstyle=\ttfamily]{#1}~}

\newtcbox{\inlinebox}[1][]{enhanced,
 box align=base,
 nobeforeafter,
 colback=blueish,
 size=small,
 left=0pt,
 right=0pt,
 boxsep=2pt,
 #1}

\renewcommand{\cref}[1]{\Cref{#1}}

\newcommand{\rom}[1]{\uppercase\expandafter{\romannumeral #1\relax}}

\newcommand{\eg}{\hbox{\emph{e.g.,}}\xspace}
\newcommand{\ie}{\hbox{\emph{i.e.,}}\xspace}

\newcounter{hypothesis}


%% file: body/0.abstract.tex
\begin{abstract}
Deep learning vulnerability detection has shown promising results in recent years. However, an important challenge that still blocks it from being very useful in practice is that the model is not robust under {\it perturbation} and it cannot generalize well over the {\it out-of-distribution (OOD)} data, e.g., applying a trained model to unseen projects in real world. We hypothesize that this is because the model learned non-robust features, e.g., variable names, that have {\it spurious correlations} with labels.  When the perturbed and OOD datasets no longer have the same spurious features, the model prediction fails. To address the challenge, in this paper, we introduced {\it causality} into deep learning vulnerability detection. Our approach \tool{} consists of two phases. First, we designed novel perturbations to discover spurious features that the model may use to make predictions. Second, we applied the {\it causal learning} algorithms, specifically, {\it do-calculus}, on top of existing deep learning models to systematically remove the use of spurious features and thus promote causal based prediction. Our results show that \tool{} consistently improved the model accuracy, robustness and OOD performance for all the state-of-the-art models and datasets we experimented. To the best of our knowledge, this is the first work that introduces {\it do calculus based causal learning} to software engineering models and shows it's indeed useful for improving the model accuracy, robustness and generalization. Our replication package is located at \url{https://figshare.com/s/0ffda320dcb96c249ef2}.


\mycomment{\ira{Current deep learning vulnerability detection models face significant challenges in the realm of generalization and robustness. Current models are not robust under perturbation and cannot generalize well over out-of-distribution (OOD) data. We hypothesize
that this is because the model learns non-robust features (e.g.,
variable names) that have spurious correlations with the labels. When
the perturbed and OOD datasets no longer have the same spurious
features, the model prediction fails. To address this challenge, we introduce causality into deep learning based vulnerability detection.
Our approach CasualVul consists of two steps. First, we design
novel perturbations to discover spurious features the model may
use to make a prediction. Second, we apply causal learning
algorithms on top of existing deep learning models to remove the
use of spurious features and promote causal-based
prediction. Our results show that we can consistently improve the model accuracy, robustness, and OOD performance for the state-of-the-art
deep learning vulnerability detection models. To the best of our
knowledge, this is the first work that introduces causal learning
and do calculus to software engineering models and shows its use
for improving model accuracy, robustness, and generalization.}
}

\mycomment{
\wei{my draft of spurious, causal features}

Causal features: the correlations of X and Y will be invariant with regard to causal features across domains, e.g., from training to testing.

spurious features: features only work for certain domains and the correlation of Y and X based on spurious features will change across domains

In the domain of vulnerability detection: a sequence of statements that determine vulnerability conditions: e.g., if we determine buffer size is smaller than string length, the buffer overflow can occur at the buffer access. The paths  that compute buffer size and string length is causal feature to buffer overflow. Variable names, function names do not contribute to the values of buffer size or string length, they are spurious features.
}
\end{abstract}

%% file: body/1.introduction.tex
\section{Introduction}
\label{sec:intro}
A source code vulnerability refers to a potential flaw in the code that may be exploited by an external attacker to compromise the security of the system.
Vulnerabilities have caused significant data and financial loss in the past~\cite{CyberCrimeLoss,MicrosoftExchangeFlaw}. Despite numerous automatic vulnerability detection tools that have been developed in the past, vulnerabilities are still prevalent. The National Vulnerability Database received and analyzed a staggering number of 16,000 vulnerabilities in the year 2023 alone~\footnote{\url{https://nvd.nist.gov/general/nvd-dashboard}}. The Cybersecurity Infrastructure Security Agency (CISA) of the United States government reported that, since 2022, there have been over 850 documented instances of known vulnerabilities being exploited in products from more than 150 companies, including major tech firms such as Google, Microsoft, Adobe, and Cisco~\footnote{\url{https://www.cisa.gov/known-exploited-vulnerabilities-catalog}}. 

Due to the recent advancements in deep learning, researchers are working on utilizing deep learning to enhance vulnerability detection capabilities and have achieved promising results. Earlier models like Devign~\cite{zhou2019devign} and ReVeal ~\cite{chakraborty2020deep} relied on architectures such as GNNs, while more-recent state-of-the-art (SOTA) models have moved towards transformer-based architectures. CodeBERT ~\cite{feng2020codebert} uses {\it masked-language-model (MLM)} objective with a replaced token detection objective on both code and comments for pretraining. GraphCodeBERT ~\cite{guo2020graphcodebert} leverages semantic-level code information such as data flow to enhance their pre-training objectives. The more recent model UniXcoder~\cite{guo2022unixcoder} leverages cross-modal contents like AST and comments to enrich code representation. 


\mycomment{
\rasel{The more recent model UniXcoder leverages cross-modal  ontents like AST and comments to enrich code representation. Almost all these transformer models have billions of parameters and they learn a different set of features during pre-training, which is reflected in down-stream tasks like vulnerability detection.} \wei{improve: rasel}

\ira{Try 2: Earlier models~\cite{zhou2019devign, chakraborty2020deep} relied on architectures such as GNNs. These approaches leverage graph representation while more recent approaches have employed statement-level vulnerability detection. More-recent models have moved towards transformer-based architectures. These transformer-based models ~\cite{feng2020codebert, guo2020graphcodebert} are trained on large text corpora and employ self-supervised objectives such as the masked-language-model (MLM) objective. The pre-trained models are then fine-tuned and can be used for vulnerability detection.  CodeBERT combines the MLM objective with a replaced token detection objective. Other approaches, like GraphCodeBERT , leverage semantic-level code information such as data flow to enhance their pre-training objectives.}}

\begin{figure}[htb]
\centering
\includegraphics[width=\columnwidth]{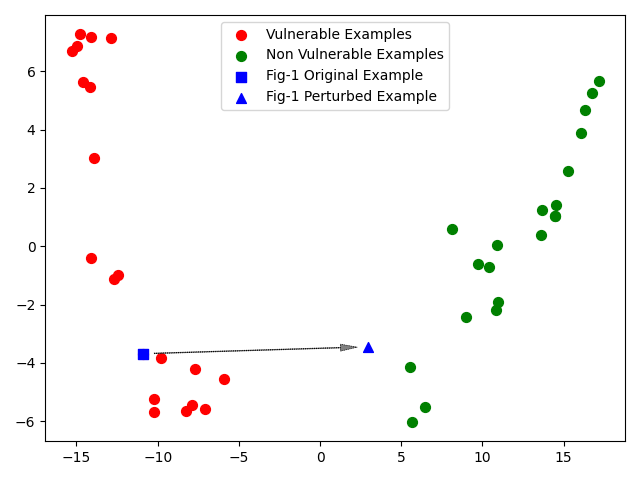}
\caption{Visualization using Principle Component Analysis
(PCA) of Figure 1’s code representations generated by CodeBERT before  and after perturbed the names.}
\label{fig:pca}
\end{figure}

However, an important challenge of deep learning tools is that the models learned and used {\it spurious correlations} between code features and labels, instead of using root causes, to predict the vulnerability. We call these features used in the spurious correlations {\it spurious features}. As an example, in \Cref{fig:vuln_ex}, this code contains a memory leak. The SOTA model CodeBERT detected this vulnerability correctly with very high confidence (probability = 0.95). However, after we refactored the code and renamed the variables (\cref{fig:perturbed_vuln_ex}), the model predicted this function as non-vulnerable. In Figure~\ref{fig:pca}, we show that the change of the variable names caused its code representation to move from vulnerable to non-vulnerable clusters.

Apparently, the model did not use the cause of the vulnerability that "the allocated memory has to be released in every path" for prediction. In this example, \linecode{av_malloc}  and \linecode{av_freep} are {\it causal features}---it is the incorrect use of \linecode{av_freep} API that leads to the vulnerability. However, the model associated {\tt s}, {\tt nbits} and {\tt inverse} with the vulnerable label  and associated {\tt out1}, {\tt dst0} and {\tt out0} with the non-vulnerable label. Such correlations are spurious; those variable names are spurious features. We believe that spurious correlations are an important reason that prevent the models from being robust and being able to apply to unseen projects.

\begin{figure*}[t]
\centering
\begin{subfigure}[t]{0.49\textwidth}
\centering
\begin{lstlisting}[mathescape=True]
FFTContext *av_fft_init(int nbits, int inverse}){
    FFTContext *s = $\textbf{av\_malloc}$(sizeof(*s));
    if (s && ff_fft_init(s, nbits, inverse))
        $\textbf{av\_freep}$(&s);
    return s;
}
// Prediction probability: 0.9493
\end{lstlisting}
\caption{Vulnerable code - Correctly Predicted}
\label{fig:vuln_ex}
\end{subfigure}\hfill
\begin{subfigure}[t]{0.49\textwidth}
\centering
\begin{lstlisting}[mathescape=True]
FFTContext *av_fft_init(int dst0, int out0){
    FFTContext *out1 = $\textbf{av\_malloc}$(sizeof(*out1));
    if (out1 && ff_fft_init(out1, dst0, out0))
        $\textbf{av\_freep}$(&out1);
    return out1;
}
// Prediction probability: 0.2270
\end{lstlisting}
\caption{Perturbed Vulnerable code - Mispredicted}
\label{fig:perturbed_vuln_ex}
\end{subfigure}
\caption{A vulnerable example predicted as vulnerable with 0.9493  but predicted as non-vulnerable with probability 0.2270 when names are perturbed by some of the spurious names from the opposite class.}
\label{fig:1}
\end{figure*}

To address this challenge, in this paper, we introduce {\it causality} (a branch of statistics) into deep learning vulnerablity detection. We developed \tool{} and applied a causal learning algorithm that implements 
{\it do calculus} and the {\it backdoor criterion} in causality, aiming to disable models to use spurious features and promote the models to use causal features for prediction.  

Our approach consists of two phases. In the first phase, we worked on discovering the spurious features a model uses. This task is challenging due to the inscrutable nature of deep learning models. To tackle this issue, we designed the novel perturbation method, drawing inspiration from adversarial robustness testing~\cite{wang2022recode}. The perturbation changes lexical tokens of code but preserves the semantics of code like compiler transformations. In particular, we hypothesized that the models may use variable names as a spurious feature like~\Cref{fig:1}, and the models may also use API names as another spurious features. Correspondingly, we designed {\it PerturbVar} and {\it PerturbAPI}, two methods to change the programs and then observe if the models' predictions for these programs are changed. Through our empirical studies, we validated that the models indeed use certain variable names and API names as spurious features, and for vulnerable and non-vulnerable examples, the models use different sets of such names  (Section \ref{sec:spurious}).


To disable the models from using such spurious features, in the second phase, we applied causal learning, which consists of special training and inference routines on top of  existing deep learning models. The causal learning puts a given model under {\it intervention}, computing {\it how the model would behave if it does not have spurious features}. This can be done via {\it backdoor criteria} from just the training data as well as our knowledge of spurious features. Specifically, we first train a model and explicitly encode a known spurious feature $F$ into the model. At the inference time for an example $x$, we use the x's representation joint with a set of different spurious features, so that the model cannot use $F$ to make the final decision for $x$.


We evaluated \tool using  Devign~\cite{zhou2019devign} and Big-Vul~\cite{BigVul}, two real-world vulnerability detection datasets and investigated on three SOTA models -- CodeBERT, GraphCodeBERT, and UnixCoder. Experimental evaluation shows that the causal model in \tool learns to ignore the spurious features, improving the overall performance on vulnerability detection by 6\% in Big-Vul and 7\% in the Devign dataset compared to the SOTA. The \tool also demonstrates significant improvement in generalization testing, improving the  performance on the Devign dataset trained model up to 100\% and Big-Vul dataset trained model up to 200\%. Experimental results also show that our \tool is more robust than the current SOTA models. It improves the performance up to 62\% on Devign and 100\% on Big-Vul on the perturbed data we constructed for robustness testing. 

\mycomment{
\begin{question}
    \ding{43} \em How robust are the DLVD models in the presence of spurious features in code?
\end{question}
}

In summary, this paper made the following contributions:
\begin{itemize}
\item We discovered and experimentally demonstrated that variable names and API names are used as spurious features in the current deep learning vulnerability detection models, 
\item We formulate deep learning vulnerability detection using causality and applied causal deep learning to remove spurious features in the models, and 
\item We experimentally demonstrated that causal deep learning can improve model accuracy, robustness and generalization. 
\end{itemize}

%% file: body/2.overview.tex
\mycomment{
\section{A Motivating Example}

intepreability tools show that the models focus on the spurious names

We currently predicted this example, and the inteprability scores changed 

Deep learning that can use root cause to determine vulnerability. For example, in Figure 1, we should use buffer overflow and its conditions at buffer write to determine the vulnerability, not the name of XXX.
}

\section{An overview of our Approach and its Novelty}
\begin{figure}[htb]
\centering
\includegraphics[width=\columnwidth]{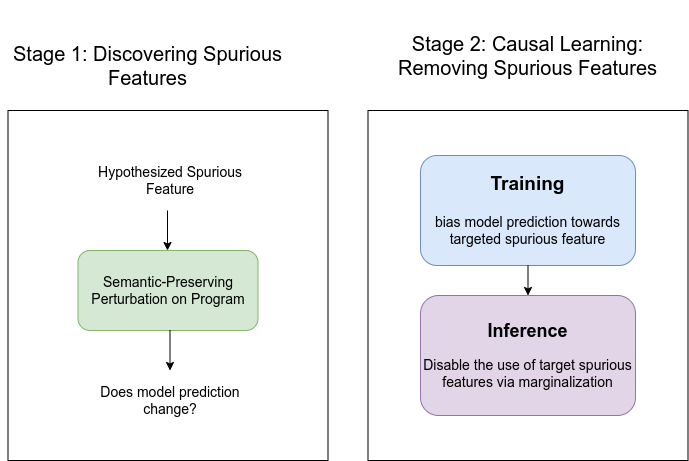}
\caption{\tool{}: an overview}
\label{fig:framework}
\end{figure}

In Figure~\ref{fig:framework}, we present an overview of our approach, \tool. 
\tool consists of two stages: (i) discovering spurious features and (ii) using causal learning to remove the spurious features. 

\paragraph{Discovering Spurious Features.} 
First, we work on discovering spurious features used in deep learning vulnerability detection models.
Spurious features are the attributes of the input dataset, \eg variable names, that exhibit correlation with the target labels (\ie vulnerable/non-vulnerable) but are not the root causes of being (non)-vulnerable, and thus, may not generalize to a desired test distribution. To discover the spurious features, in our approach, 
we first hypothesize a spurious feature based on our domain knowledge.
For instance, we assumed variable names can be spurious features for vulnerability detection tasks---models should not make a decision based on the names.
Next, we design {\it perturbations} based on each hypothesized feature. Specifically, we transform the programs via a refactoring tool that changes the feature, but does not change the semantics of code (similar to compiler transformations). After we observe the model changes its prediction for many such examples, we conclude that the model likely relied on this spurious feature. Discovering spurious features is a very challenging task. 
As an instance, randomly altering variable names does not reveal its spurious nature. 
We need to carefully identify which variable(s) need to be changed by which name(s) to maximize its impact.
To the best of our knowledge, we have not seen a systematic study on finding spurious features in vulnerability detection models.

\paragraph{Removing Spurious Features.} Once we discovered spurious features, in this step we try to reduce the impact of these features on the model decision. One easy way to achieve that could be to
augment the training data by randomizing the spurious features and then retrain the model. However, it will take much time for big, pretrained, code representations/deep models. So instead, we take existing representation and apply causal learning to disable the model to use spurious features at inference time. This ensures the model can rely mostly on the causal features for prediction. 



We accomplish this goal as follows. 
First, as an input, we take the existing representation and the known spurious feature to be eliminated.
We then train a model such that the representation learned in this step is especially biased towards the targeted spurious feature.
Next, during inference, the model makes a decision based on the input representation while ignoring the bias representation learned in the previous step. 
In particular, we marginalized over a set of examples that contain different spurious features and prevent the inference from utilizing the targeted spurious feature while making the final decision. 
This approach relies on the principles of "do-calculus" and the "backdoor criterion," which we will explain in detail in Sections~\ref{sec:do} and~\ref{sec:alg}, respectively.

To the best of our knowledge, this is the first work to bridge causal learning with deep learning models for vulnerability detection.
It helps researchers understand to what degree causality can help deep learning models remove spurious features and to what degree we can improve model accuracy, robustness and generalization after removing spurious features.
Such causal learning approaches have been applied in computer vision~\cite{2022icsecito}. A key difference is that the vision domain is {\it continuous} in that the pixel values are continuous real numbers, while program features are discrete. Thus, in the domain of program data, we can explicitly discover spurious features and then apply causal learning in a targeted manner.


%% file: body/3.spuriousfeatures.tex
\newcommand{\var}{\textit{PerturbVar}\xspace}
\newcommand{\dead}{\textit{PerturbAPI}\xspace}
\newcommand{\joint}{\textit{PerturbJoint}\xspace}

\section{Discovering Spurious Features}
\label{sec:spurious}





In this section, we illustrate our technique for discovering spurious features through semantic-preserving perturbations. 

Spurious features are features that spuriously correlate to the output. 
Those spurious correlations often only exist in training but not in testing, especially testing under perturbed data or unseen projects. That is because a feature that is spurious, as opposed to causal, will change across domains and is no longer useful for prediction, as shown in Figure~\ref{fig:1}. Thus, a standard way to determine whether one feature is spurious is to \textit{perturb} the values of the feature and observe how the output changes.  


Following this idea, we first hypothesized two spurious features, {\it variable names} and {\it API names} that the current deep learning models may use for vulnerability detection. 
We chose these two spurious features as a proof of concept to check if we perturb in 
lexical and syntactic characteristics respectively while keeping the overall semantic of the program the same, whether the model prediction would change. We applied an existing code refactoring tool, {\it NatGen}~\cite{data_avail}, to perturb the code in test dataset, and observe whether the model performance is changed between the original test set and the perturbed test set (see problem formulation in Section~\ref{problem}). Interestingly, we found that randomly changing the variable names and function names do not bring down the model performance. Thus we designed novel perturbation methods that can demonstrate the two spurious features. See Sections~\ref{sec:var} to~\ref{sec:join}.






\subsection{Problem Formulation}~\label{problem}
Given a perturbation $p$, a code sample $s$, and a trained model $M$, we discover a spurious feature if the following conditions are met: (1) the application of the perturbation, $p(s)$, does not alter the semantics of the function, (2) the model's prediction changes upon transformation, and (3) the candidates for a perturbation $p$ are drawn from the training distribution.

\mycomment{
\begin{enumerate}
    \item The application of the perturbation, $p(s)$ does not alter the semantics of the function, like compiler transformation.
    \item The model's prediction changes upon transformation, \ie the model's prediction should flip upon the application of the perturbation. That is $M(s)$ $\neq$ $M(p(s))$
    \item The candidates for a perturbation $p$ are drawn from the training distribution
\end{enumerate}
}

We will use F1 score as a more-thorough evaluation metric for our imbalanced datasets and condition number (2). A degraded F1 score indicates the application of the perturbation $p(s)$ has resulted in mis-classifications (flipped labels).

 
\mycomment{
We achieve this with three core strategies:
\begin{itemize}
    \item \textbf{Perturbation in the lexical space}: We perform perturbations on existing variable names. We replace existing variable names with names from the training set that are controlled by frequency in the opposite class label.
    \item \textbf{Perturbation in the syntactic space}: We perform perturbations using dead-code insertion. We construct unreachable code using API calls from the training data. API calls are controlled by frequency in the opposite class label. 
    \item \textbf{Perturbation in both the lexical \& syntactic space}: We perform perturbation in a joint setting, by composing the former perturbations.
\end{itemize}

We evaluate test-set performance \textit{with} and \textit{without} the perturbations applied. A change in performance for the same test dataset indicates a spurious correlation. If there is an observed degradation in performance across domains, then we confirm the hypothesized spurious feature. 

We assess the effectiveness of our spurious feature discovery on 3 model types: CodeBERT, GraphCodeBERT, and UniXcoder. Furthermore, we ensure to respect the following constraints:

\textbf{Consistency.} We ensure that multiple occurrences of the same identifier are replaced consistently in a given code sample. That is, if a variable $VAR_1$ appears in multiple locations, we ensure it is replaced with the same spurious variable name consistently throughout the sample. 

\textbf{Naturalness.} Feature candidates are drawn from the training distribution. Both variable names and API calls are drawn from the training distribution and ranked by frequency. This setup allows us to confirm strong spurious correlations. 

\textbf{Semantic preservation} Both variable swap and dead-code insertion are semantic-preserving transformations. By definition, dead code is a block of code that is not executed in the sample. By adhering to these strict conditions, we ensure the meaning of the input code is not altered.
}

\subsection{\var: Variable Name as a Spurious Feature}~\label{sec:var} 
To demonstrate that some variable names are indeed used by the models as spurious features, we design an extremely perturbed test set. We analyze the training dataset and sort the variable names based on their frequency of occurrences in vulnerable functions and non-vulnerable functions, respectively. When replacing existing variable names in the test set, we randomly select a name from the top-K most-frequent variable names of the {\it opposite-class labels}: for a non-vulnerable sample, we replace the existing variable names with a name from the vulnerable training set, but which does not occur in the non-vulnerable training set, and vice-versa. We apply this perturbation to every sample in the test set.



\begin{table}[h]
\centering
\caption{PerturbVar: Impact of Variable Name Perturbation}~\label{tab:var} 
\resizebox{\columnwidth}{!}{
\begin{tabular}{c|cc|cc|cc}

\toprule
{\multirow{3}{*}{\begin{tabular}[c]{@{}c@{}}Top-K \\ (Freq.)\end{tabular}}} 
& \multicolumn{2}{c|}{\textbf{CodeBERT}}                                       
& \multicolumn{2}{c|}{\textbf{UniXcoder}}                                      
& \multicolumn{2}{c}{\textbf{GraphCodeBERT}}  \\ \cline{2-7}

 & Devign & Big-Vul & Devign & Big-Vul & Devign & Big-Vul   \\
 
& \textbf{F1} & \textbf{F1} & \textbf{F1} & \textbf{F1} & {\textbf{F1}} & {\textbf{F1}} \\ \midrule
{Baseline}    & {0.61}      & {0.36}  & {0.63}  & {0.38}  & {0.62}  & {0.37}   \\ \midrule
{Random}      & {0.61}      & {0.32}  & {0.63}  & {0.36}  & {0.62}  & {0.35}   \\ \midrule
{Top 100}     & {0.61}      & {0.25}  & {0.52}  & {0.24}  & {0.60}  & {0.27}   \\
{Top 50}      & {0.55}      & {0.25}  & {0.55}  & {0.22}  & {0.59}  & {0.26}   \\
{Top 25}      & {0.54}      & {0.26}  & {0.56}  & {0.25}  & {0.59}  & {0.27}   \\
{Top 20}      & {0.55}      & {0.25}  & {0.56}  & {0.24}  & {0.59}  & {0.28}   \\
{Top 15}      & {0.54}      & {0.23}  & {0.56}  & {0.24}  & {0.59}  & {0.27}   \\
{Top 10}      & {0.54}      & {0.26}  & {0.53}  & {0.23}  & {\textbf{0.57}}   & {0.28}   \\
Top 5         & \textbf{0.52}  & {\textbf{0.21}}  & \textbf{0.52}   & \textbf{0.18}  & 0.58 & \textbf{0.26} \\ \bottomrule                                                                      
\end{tabular}
}
\end{table}

\textbf{Observation.}~As shown in Table~\ref{tab:var}, we are able to degrade the F1 score as much as 11.5 \% on the Devign dataset, and as much as 20 \% on the Big-Vul dataset. We have observed the performance degradation across all the datasets and  multiple architectures: CodeBERT, GraphCodeBERT, and UniXcoder.  
However, in the randomized setting, the performance almost does not change relative to the baseline.  Introducing common vulnerable names into non-vulnerable code-samples causes the model to misclassify the sample as vulnerable. Conversely, introducing common non-vulnerable names into vulnerable code samples
causes the model to misclassify the sample as non-vulnerable.  The more common the variable names are used (i.e. the lower the Top-K), the more the performance degrades.

\subsection{\dead: API Name as a Spurious Feature}
Modern programs frequently use API calls. We conjecture that the models may establish spurious correlations between API names with vulnerabilities. Similar to the approach used in \var, we ranked the frequency of the API calls in the training data, for vulnerable examples and non-vulnerable examples respectively. We then insert API calls (that are ranked in the top-100 occurring calls in non-vulnerable examples, but which are not frequently-occurring in the vulnerable examples) into vulnerable examples and vice-versa. To preserve the semantics of code, we insert these API calls as "dead-code", \ie this code will never be executed.


 
 We inject dead-code at $n$ random positions within the code sample. The dead-code block is composed of $m$ distinct function/API calls ($m$ and $n$ are configurable and we used $m=5, n=5$ to produce the results in Table~\ref{tab:api}). We guard the block of API-calls with an unsatisfied condition, ensuring the loop is never executed, as shown in Figure \ref{fig:deadcodefig}: 

 \begin{figure}[!h]
     \centering
     \begin{lstlisting}[language=C, basicstyle=\scriptsize \ttfamily, numbers=left]
    while ( _i_4 > _i_4 ) { 
    tcg_out_r(s , args[1]); help_cmd(argv[0]); 
    cris_alu(dc , CC_OP_BOUND , cpu_R[dc -> op2] , cpu_R[dc -> op2] , l0 , 4);
    RET_STOP(ctx); tcg_out8(s , args[3]); }
    \end{lstlisting}
        \caption{Dead-code composed of our spurious feature, API calls}
     \label{fig:deadcodefig}
 \end{figure}




\begin{table}[h]
\centering
\caption{PerturbAPI: Impact of API Name Perturbation} 
~\label{tab:api}
\resizebox{\columnwidth}{!}{
\begin{tabular}{c|cc|cc|cc}

\toprule
{\multirow{3}{*}{\begin{tabular}[c]{@{}c@{}}Dead-code \\ Type\end{tabular}}} 
& \multicolumn{2}{c|}{\textbf{CodeBERT}}                                       
& \multicolumn{2}{c|}{\textbf{UniXcoder}}                                      
& \multicolumn{2}{c}{\textbf{GraphCodeBERT}}  \\ \cline{2-7}

 & Devign & Big-Vul & Devign & Big-Vul & Devign & Big-Vul   \\
 
& \textbf{F1} & \textbf{F1} & \textbf{F1} & \textbf{F1} & {\textbf{F1}} & {\textbf{F1}} \\ \midrule
{Baseline}    & {0.61}      & {0.36}  & {0.63}  & {0.38}  & {0.62}  & {0.37}   \\ \midrule
{Random}      & {0.61}      & {0.35}  & {0.62}  & {0.36}  & {0.62}  & {0.35}   \\ \midrule
API         & \textbf{0.52}  & {\textbf{0.10}}  & \textbf{0.34}   & \textbf{0.11}  & \textbf{0.47} & \textbf{0.09} \\ \bottomrule                                                                       
\end{tabular}
}
\end{table}

\textbf{Observation.} We compare our results to (1) baseline performance and (2) random dead-code transformation performance, shown in Table~\ref{tab:api}. Our results show that dead-code composed of API calls severely hurts model performance compared to the vanilla baseline and random dead-code transformation. Model performance degrades proportionally with the inclusion of increased API calls and increased injection locations. Performance in the vanilla model degrades as much as 28.73 \% on Devign and 27.99 \% on Big-Vul.

\subsection{\joint: Combine Them Together}\label{sec:join} 
We hypothesize that the composition of the two spurious features will further degrade model performance. We setup the study, where for every sample constructed in the 
\dead dataset from Section~\ref{tab:api}, we replace existing variable names with a random selection from the top-K most-frequent variable names of the opposite class (using the same approach from Section~\ref{tab:var}). 

\textbf{Observation.}~In Table~\ref{tab:joint}, our results show that the composition of API dead-code and variable renaming further degrades the model. The model degradation is more severe when applying the composition of the settings. In the combined setting, performance degrades as much as 41.37 \% for Devign and 33.89 \% for Big-Vul.



\begin{table}[h]
\centering
\caption{PerturbJoint: Impact of Joint Perturbation}
~\label{tab:joint}
\resizebox{\columnwidth}{!}{
\begin{tabular}{c|cc|cc|cc}

\toprule
{\multirow{3}{*}{\begin{tabular}[c]{@{}c@{}}Top-K \\ (Freq.)\end{tabular}}} 
& \multicolumn{2}{c|}{\textbf{CodeBERT}}                                       
& \multicolumn{2}{c|}{\textbf{UniXcoder}}                                      
& \multicolumn{2}{c}{\textbf{GraphCodeBERT}}  \\ \cline{2-7}

 & Devign & Big-Vul & Devign & Big-Vul & Devign & Big-Vul   \\
 
& \textbf{F1} & \textbf{F1} & \textbf{F1} & \textbf{F1} & {\textbf{F1}} & {\textbf{F1}} \\ \midrule
{Baseline}    & {0.61}      & {0.36}  & {0.63}  & {0.38}  & {0.62}  & {0.37}   \\ \midrule
{Top 100}     & {0.38}      & {0.07}  & {0.23}  & {0.07}  & {0.59}  & \textbf{0.06}   \\
{Top 50}      & {0.37}      & {0.07}  & {0.25}  & {0.06}  & {0.60}  & {0.07}   \\
{Top 25}      & {0.36}      & {0.07}  & {0.28}  & {0.08}  & {0.60}  & {0.07}   \\
{Top 20}      & {0.38}      & {0.08}  & {0.27}  & {0.08}  & {0.48}  & {0.07}   \\
{Top 15}      & {0.36}      & {0.07}  & {0.27}  & {0.06}  & {0.52}  & {0.07}   \\
{Top 10}      & {0.37}      & {0.08}  & {0.24}  & {0.07}  & {\textbf{0.45}}   & {0.08}   \\
Top 5         & \textbf{0.33}  & {\textbf{0.06}}  & \textbf{0.22}   & \textbf{0.05}  & 0.47 & 0.07 \\ \bottomrule                                                                       
\end{tabular}
}
\end{table}



\paragraph{\bf Summary.} In this section, we investigated multiple datasets and multiple SOTA models, revealing that variable names and API names are spurious features frequently utilized by these models. Interestingly, the models associate different variable names and API names as spurious features for different labels. 
Consequently, it became evident that only meticulously designed perturbations, not random ones, showcased the usage of these spurious features, consequently leading to a decline in the models' performance on the perturbed datasets.

%% file: body/3.causality.tex
\section{Causal Learning to Remove Spurious Features}~\label{sec:do}
\label{sec:background}


In this section, we present how to apply causal learning to remove spurious features in the vulnerability models.




\subsection{Causal Graph for Vulnerability Detection}
\begin{figure}[thb]
\centering
\includegraphics[width=\columnwidth]{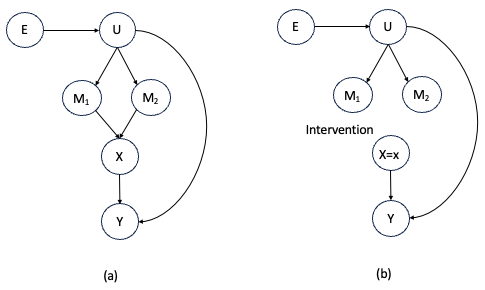}
\caption{Causal Graph Before and After Do Calculus }
\label{fig:causalstructure}
\end{figure}

To apply causal learning, our first step is to construct a {\it causal graph} to model how vulnerability data is generated from a statistics point of view, shown in Figure~\ref{fig:causalstructure} (a).
A causal graph visually represents the causal relationships between different variables or events. In the graph, nodes represent random variables, and directed edges between nodes indicate causal relationships. Thus, A$\rightarrow$B means variable A directly influences variable B. 
The absence of an edge between two nodes implies no direct causal relationship between them. 
In Figure~\ref{fig:causalstructure} (a), $X$ represents a function in the program. Whether there is a vulnerability or not in the code is directly dependent on the function. 
So we add an edge from $X$ to $Y$, where $Y$ is the label of vulnerability detection, 1 indicates vulnerable, and 0 indicates not vulnerable.



In the causal graph, we use $E$, namely {\it environment}, to model the domain where the dataset is generated. For example, at training time, namely $E = 0$, this environment indicates e.g., the code is written by certain developers and for certain software applications. At testing or deployment time of the model, namely $E = 1$, the code may link to different developers and applications. Such environment-specific factors are modeled using $U$. It is a {\it latent} variable and not directly observed in the training data distribution. For example, $U$ can denote the expertise 
or coding style of the programmer or the type of software application. Because of $U$, there can be different spurious features in the code, like the two we showed in Section~\ref{sec:spurious}, denoted as $M_1$ and $M_2$. 

\mycomment{
In Figure~\ref{fig:causalstructure} (a), 
$U$ has a path to both $X$ \cz{U does not point to X directly, we need to explain how U creates M, then how M creates X.} and $Y$, representing that it can impact both code and its vulnerability. For example, $U$ can denote the expertise 
or coding style of the programmer or the type of software application. 
}

$M_1$ represents variable naming styles, as developers may like to use certain formats of variable names in their code. Similarly, $M_2$ represents API names, as certain developers or applications may more likely use a particular set of API calls. These basic (text) features of $M_1$ and $M_2$ influence the code. Thus, we add the edges $X$ to $M_1$ and $M_2$, respectively. Note that the causal graph can be expanded to integrate more of such spurious features. We leave this to future work. 

In Figure~\ref{fig:causalstructure} (a), we also have the edge $U$ to $Y$,  indicating that $U$ can impact $Y$. For example, a junior developer may more likely introduce vulnerabilities; similarly, certain APIs may more likely introduce vulnerabilities, e.g., SQL injection. 

\subsection{Applying Causality}
When we train a deep learning model, using the code $X$ as input, to predict the vulnerability label $Y$, we learn the correlation $P(Y|X)$ in the training dataset.  However, when we deploy this trained model in a new environment, e.g., handling perturbed datasets or unseen datasets., the domain will be changed. Due to the shifting of $E$ from 0 to 1, the correlation $P(Y|X)$ will be different, denoted as $P(Y|X, E=0) \neq P(Y|X, E=1)$. From the causal graph point of view, $X$ and $Y$ are both the descendants of $E$ and therefore will be affected by its change. 

To improve the models' performance in generalizing to new environments beyond the training, our goal is to learn the signal that is invariant across new environment and domains. Applying causal learning, we {\it intervene} on the causal graph so that the correlation between $X$ and $Y$ is the same no matter how the environment $E$ changes. In causality, such intervention is performed via {\it do-calculus}, denoted as $P(Y|do(X=x))$. 

The key idea is that we imagine to have an oracle that can write perfect, unbiased code $x$; we then use such code to replace the original code $X$ for vulnerability detection. Doing so, we created a new causal graph by removing all the incoming edges to the node $X$ and put the new code here $X=x$. See Figure~\ref{fig:causalstructure} (b). In this new causal graph, $E$ only can affect $Y$, not $X$, and thus the correlation of $X$ and $Y$ will no longer be dependent on $E$. In other words, the correlation of $x$ and $Y$ in the new causal graph then becomes the \emph{invariant} signal across different  domains, e.g., from training to testing. When we learn the correlation in this new causal graph, the learned model can generalize to new environments since this relationship also holds at testing. In the next section, we explain how we compute $P(Y|do(X=x)$ from the observational data.

\subsection{Estimating Causality through Observational Data}
Figure~\ref{fig:causalstructure} (a) models the joint distribution of vulnerability data from a data generation point of view. Figure~\ref{fig:causalstructure} (b) presents the causal graph after performing do calculus on $X$, allowing us to generalize across the domains. The challenge is that performing intervention and obtaining the perfect, unbiased code is almost impossible. To estimate the causal effect without actually perform the intervention, we apply the {\it backdoor criterion}~\cite{causalbook} to
derive causal effect $P(Y|do(X))$ from the given observational data of $X$ and $Y$. 

Backdoor criterion teaches how to block all the non-causal paths from $X$ to $Y$ in the causal graph so we can compute the causal effect. For example, in Figure~\ref{fig:causalstructure} (a), node $U$ can cause $X$, $U$ can cause $Y$; as a result,  there can exist correlations between $X$ and $Y$ in the observational data. However, such correlation do not necessarily indicate the causal relation from $X$ to $Y$. For example, the junior developers may like to write their code with the variable name {\tt myvar} and the junior developers may more likely introduce the vulnerability into their code. In this case, we might often see the code with {\tt myvar} is detected with vulnerability in the dataset. However, {\tt myvar} is not the cause for the vulnerability. The correlation of {\tt myvar} and vulnerability is spurious. To remove such spurious correlations, the backdoor criterion states that we need to condition $X$ on $M_1$ and $M_2$; as a result, we can remove the incoming edges to $X$ and thus impact from $U$ to $X$. Mathematically, we will have the following formula for Figure~\ref{fig:causalstructure} (b).

\begin{equation}
    P(Y|do(X=x)) = \sum_{M_1, M_2} P(Y|X=x, M_1, M_2) P(M_1)P(M_2)
\end{equation}




To compute Eq.(1), we adopted an algorithm from ~\cite{2022icsecito}. First, we train a model that predicts $Y$ from $R$ and $M$ (we denote $M_1, M_2$ as $M$ for simplicity), denoted as $P(Y|R,M)$, where $R$ is the  representation of $X$ computed by an existing deep learning model, and $M$ is the representation of $X$ that encodes the spurious features. For example, if we encode each token of $X$ into $M$, we will have variable names and API names encoded in $M$. Since $R$ is a representation learned by models, the algorithm assumes $R$ encoded both causal and spurious features. The goal of taking $R$ and $M$ jointly for training is to especially encode targeted spurious feature(s) in the $M$ component of the model $P(Y|R,M)$.




At inference, we use the model $P(Y|R,M)$ to compute $P(Y|do(X=x))$ via Eq. (1). The backdoor criterion instructs that we first randomly sample examples from $X$ that contains different spurious features. We will then marginalize over the spurious features through weighted averaging, that is, $\sum_{M_1, M_2} P(Y|X=x, M_1, M_2) P(M_1, M_2)$. This can be computed using the model trained above $P(Y|R,M)$.
Intuitively, the spurious features in the data have been cancelled out due to the weighted averaging (please refer to ~\cite{pearl2011transportability} to understand why this is the case). The model will make predictions based on the remaining signals, which are the causal features left in $R$.

\mycomment{Learn casual effects, we need to perform {\it do} calculus. Intervention that estimate the probability based on only cause $C$ of the vulnerability  

$P(Y|X)$ = causal + spurious 

$P*(Y|X)$ = causal + spurious*

$P(Y|X)\neq P*(Y|X)$ 

$P(Y|do(X)) = P*(Y|do(X))$ \wei{should we use C or X here}

$P(Y|do(X))$ use causal features only to predict Y

Difference:
Here, we assume that not all the back door variables are observable 

\subsection{Causal Diagram and Backdoor Variables}
Applying causal inference, we first construct casual structure to model the vulnerability detection

\wei{is this diagram correct?}

\subsection{Computing Do Calculus: the Math Formula and the Assumptions}

Formula 1 in~\cite{2022cvprmao}

How to perform do calculus? The assumptions of our approaches

Assumption 1~\cite{2022cvprmao} applies here:
X = W (spurious) +Z (causal);
R = W (spurious) +Z (causal)

W are not confounding with Z, as the two have no direct relations?

Assumption 2~\cite{2022cvprmao} is a strong assumption, but we still assume: representation captures the causal features 

{\it require a strong model that can learn the best representation}

Assumption 3~\cite{2022cvprmao}

Step 1: build a model that uses both spurious and causal features to predict

Step 2: during inference, only can use causal features 
}

%% file: body/3.alg.tex
\section{The algorithms of Causal Vulnerability Detection}~\label{sec:alg}
In this section, we present the algorithms that compute the terms Eq. (1). In Algorithm~\ref{alg:one}, we will train a model of $P(Y|R,M)$. In Algorithm~\ref{alg:two}, we show how to apply $P(Y|R,M)$ and use backdoor criteria to undo the spurious features during inference. 

\subsection{Training P(Y|R,M)}
Algorithm~\ref{alg:one} takes as input the training dataset $D$ as well as the spurious feature(s) we aim to remove, namely {\it targeted spurious feature(s)~$t$}. For example, it can be variable names and/or API names, as presented in Section 3.

The goal of Algorithm \ref{alg:one} is to learn $P(Y|R, M)$ from the training data.  We use the embedding ($r$) of the original input $x$ computed by an existing deep learning model. At line 4, the training iterates through $n$ epochs. For each labeled data point $(x, y)$, we first obtain $r$ of $x$ at line~6. 



At line~7, we select x' that shares the targeted spurious feature $t$ with x but differ in root cause $r$. This means (1) $x'$ should have the same label as $x$; as shown in Section 3, spurious features are label specific; (2) $x'$ should share $t$ with x, so that the model will more likely encode $t$ in the $M$ component of $P(Y|R,M)$; and (3) $x'$ should not be x so that training with $(r,M_{x'}, y)$ at line~9, the model will rely on the spurious feature in $M_{x'}$ instead of $r$ to make prediction. Our final goal is to encode in the model the root cause of $x$ in the $R$ component, and the targeted spurious feature $t$ shared across x and $x'$ in the $M$ component.

\mycomment{
\begin{enumerate}[leftmargin=*]
\item $x'$ should have the same label as $x$ so that $x$ and $x'$ likely share the root cause in $r$;
\item $x'$ should share the targeted spurious feature $t$ with x; in this way, the model will more likely encode $t$ in the $M$ component of $P(Y|R,M)$; 
\item $x'$ should not be x so that training with $(r,M_{x'}, y)$ we therefore encoded in the model the root cause as well as the targeted spurious feature shared across x and $x'$, but not others.
\end{enumerate}
}




In Table~\ref{tab:alg}, we designed different ways of selecting x' at line~7, {\it Settings Var1}-{\it Var2} for removing the spurious feature of variable name, {\it Settings API1}-{\it API3} for API names, {\it Setting Var+API} is for both variable and API names.

The first idea is to select x' from the training dataset such that it shares the maximum number of {\it spurious} variable or API names with x. See the rows of {\it Var1} and {\it API1}.  The second idea is to construct an example x' such that x and x' share $k$ variable/API names. See the rows of {\it Var2} and {\it API2}. Similarly, Setting {\it API3} constructs x' to have the top $k$ spurious API names. We have also tried this approach for variable names, but it does not report good results. In Setting {\it Var+API} for removing both spurious features, we take x' from Setting {\it Var1} and then perform transformation using Setting {\it API3}. This achieved the best results in our experiments compared to other combinations. 

\begin{table*}[t]
\caption{Selection Procedures for x' in Algorithm.~\ref{alg:one}}~\label{tab:alg}
\begin{tabular}{ c | c | l }
\toprule
Spurious Feature & Setting &  x': same label of x \\\midrule
\multirow{2}{*}{Variable name} &Var1& 
select x' which shares the maximum number of spurious variable names with $x$.\\\cline{2-3}
&Var2& 
random select x', replace at most $k$ variable names of x' with the variable names from x\\\midrule
\multirow{ 4}{*}{API name}& API1&
select $x'$ which shares the maximum number of spurious API names with $x$
\\\cline{2-3}
&API2 & random select $x'$, randomly select $k$ APIs from $x$ and insert them in $x'$ as dead code \\\cline{2-3}
&\multirow{2}{*}{API3} & random select x', pick $k$ APIs from the top 10\% most frequent spurious APIs and insert them\\
&&in x' as deadcode\\
\midrule
Variable and API names & Var+API& select $x'$ based on setting {\it Var1}; then insert dead code according to setting {\it API3}.
\\\bottomrule
\end{tabular}
\end{table*}





\mycomment{The components: $P(R|X)$ -- Assumption 2
We are using the STOA models to learn a source code representation

$\hat{P}(Y|R,X')$ -- Assumption 3

X' should encode S1-Sn token level (Variable name, API names) features

Phrase 1: learn the causal as well as the spurious features

find an example that can expose spurious feature

Phrase 2:  marginalizing over the spurious features, discard the spurious models 
}

\begin{algorithm}

\caption{Causal Learning Model Training}\label{alg:one}

\textbf{Input:} Training dataset D over {(X, Y)}; Targeted spurious feature $t$


\textbf{Phase 1:} Compute $\hat{P}(R|X)$ by the SOTA code representation model

\textbf{Phase 2:} {

    \For{$i \gets 1$ \KwTo $n$ {\it epochs}}{
        \ForEach{$(x, y)$ \textbf{in} $D$}{
          Extract r for x using $\hat{P}(R|X)$.
          
         Select $x'$ by one of the {\it selection procedures} in Table~\ref{tab:alg}
          
          Encoding $x'$ to $M_{x'}$



          Train ${P}(Y|R,M)$ using $(r, M_{x'}, y)$ via minimizing the classification loss
        }
    }
}
\textbf{Output}: Model $\hat{P}(R|X)$ and ${P}(Y|R,M)$

\end{algorithm}


\subsection{Undo Spurious Features during Inference}
In Algorithm \ref{alg:two}, we explain our inference procedure. To predict the label for a function $x$, we first extract $r$ at line~2. Existing work directly predict the output $y$ using $r$. Since there are both spurious features and core features in $r$, both of them are going to be used to predict the vulnerability. Here, we will use our causal algorithm to remove the spurious features in our inference. The key intuition is to cancel out the contribution of the spurious features by averaging over the prediction from all kinds of spurious features. To do so, at line~4, we randomly sample different x' $K$ times from the training dataset D. At line~8, we then compute Eq (1). We assumed uniform distribution for $P(M_x')$. Finally, at line~9, we make the prediction. $argmax\_y$ means select the label that has the better probability among the vulnerable/nonvulnerable classes.

\begin{algorithm}

\caption{Causal Learning Model Inference}
\label{alg:two}

\textbf{Input:} Query x, training dataset D over {(X, Y)}, models $\hat{P}(R|X)$ and $\hat{P}(Y|R, M)$ 

Extract r for x using $\hat{P}(R|X)$.



\For{$i \gets 1$ \KwTo $K$}{

  Randomly select $x'$ from the training set D.

  Extract spurious features $M_{x'}$ from $x'$.
  
  Compute $\hat{P}(y_i|r, M_{x'_i})$
}

Calculate the causal effect $P(y|do(X=x))=$ $\hat \sum_i{\hat P(y_i|r, M_{x'_i})P(M_{x'_i})}$


\textbf{Output}: Class $\hat{y} = argmax_yP (y|do(X=x))$.

\end{algorithm}








%% file: body/4.experimental_setup.tex
\section{Experimental Setup}
\label{sec:experiments}


\subsection{Implementation} 
We use Pytorch 2.0.1\footnote{https://github.com/pytorch/pytorch} with Cuda version 12.1 and transformers\footnote{https://github.com/huggingface/transformers} library to implement our method. All the models are fine-tuned on single NVIDIA RTX A6000 GPU, Intel(R) Xeon(R) W-2255 CPU with 64GB ram. The pretrained weights and tokenizers of the transformer models were obtained from the link provided by the original authors~\footnote{https://github.com/microsoft/CodeBERT}.   We used Adam Optimizer to fine-tune our models. The models were trained until 10 epochs while the batch size is set to 32. The learning rate is set to 2e-5. We trained the models with our training data and the best fine-tuned weight is selected based on the f1 score against the validation set. We used this weight during the evaluation of our test set. We set K=40 (Algorithm ~\ref{alg:two} line 3), as it reported the best results among the values we tried.




\subsection{Datasets and Models} We considered two vulnerability detection datasets: Devign \cite{zhou2019devign} and Big-Vul \cite{BigVul}. Devign is a balanced dataset consisting of 27,318 examples (53.22\% vulnerable) collected from two different large C programming-based projects: Qemu and FFmpeg. As Zhou et al.~\cite{zhou2019devign} did not provide any train, test, and validation split, we used the split published by CodeXGLUE authors~\cite{CodeXGLUE}. Big-Vul dataset is an imbalanced dataset consisting of 188,636 examples (5.78\% vulnerable)  collected by crawling the Common Vulnerabilities and Exposures (CVE) database. For this dataset, we used the partitions published by the LineVul authors \cite{LineVul}. 

We evaluated the three SOTA models, CodeBERT, GraphCodeBERT and UniXcoder as the model $P(R|X)$ in Algorithm~\ref{alg:one}. The representation R is extracted from the output embedding of the last hidden layer of these models. To construct the network $\hat{P}(Y|R, M)$, at first,  we pass $x'$ through the first four encoder block of these transformer models. The obtained output embedding is considered as M. We used the first fourth encoder block (empirically it is the best layer we found) from the twelve blocks to compute M because the early layers learn the low-level features and spurious features tend to be the low-level features~\cite{2022cvprmao}. 
M is then concatenated with $R$. Finally, a 2-layer fully-connected network is used to predict $Y$.

%% file: body/5.results.tex
\section{Evaluation}
\label{sec:results}
We studied the following research questions:
\mycomment{
\begin{itemize}
\item {\bf RQ1:} Can \tool{} improve the accuracy of the model in comparison
with other state-of-the-art models?
\item {\bf RQ2. Robustness \& Generalization.} Can \tool{} improve the robustness and generalization of the model in comparison with other state-of-the-art models?
\item {\bf RQ3. Impact of Design Choice.} How do different design choices affect the performance of \tool{}?
\end{itemize}
}

\noindent{\bf RQ1:} Can \tool{} improve the accuracy of the model?

\noindent{\bf RQ2:} Can \tool{} improve the robustness and generalization of the model?

\noindent{\bf RQ3:} (ablation studies) How do different design choices affect the performance of \tool{}?

\subsection{RQ1: Model Accuracy}


\noindent
\textbf{Experimental Design.} To answer this RQ, we implemented our causal approach on top of three state-of-the-art transformer-based vulnerability detection models - CodeBERT, GraphCodeBERT, and UniXcoder. We used the default (w/o causal) versions of these models as baselines. We address these default versions as {\it Vanilla models}.


We experimented with all the causal settings shown in Table~\ref{tab:alg}, namely {\it Var1} and {\it Var2}, {\it API1}, {\it API2} and {\it API3}, and {\it Var+API}. We evaluated both the vanilla models and the causal models on the same \textit{unperturbed} original test set and use the metrics of {\it F1}.  We trained all the models three times with different random seeds and consider the average F1 score as our final score (this is done for all the RQs).  


\begin{table}[t]
\setlength{\tabcolsep}{2pt}
\caption{ The F1 score of \tool{} and the vanilla model on the test set of Devign and Big-Vul dataset. }
\begin{tabular}{l|cc|cc|cc}
\toprule
Settings  & \multicolumn{2}{c|}{CodeBERT}        & \multicolumn{2}{c|}{GraphCodeBERT}   & \multicolumn{2}{c}{UniXcoder}       \\ \midrule
          & \multicolumn{1}{c|}{Devign} & Big-Vul    & \multicolumn{1}{c|}{Devign} & Big-Vul    & \multicolumn{1}{c|}{Devign} & Big-Vul    \\ \midrule
Vanilla   & \multicolumn{1}{c|}{0.61} & 0.38 & \multicolumn{1}{c|}{0.63} & 0.38 & \multicolumn{1}{c|}{0.64} & 0.39 \\ \midrule
\multicolumn{7}{c}{\tool{}} \\ \midrule
Var1      & \multicolumn{1}{c|}{0.65} & 0.40 & \multicolumn{1}{c|}{0.63} & 0.38 & \multicolumn{1}{c|}{0.66} & 0.40 \\ 
Var2      & \multicolumn{1}{c|}{0.64} & 0.40 & \multicolumn{1}{c|}{0.63} & 0.40 & \multicolumn{1}{c|}{0.65} & \textbf{0.41} \\ 
\midrule
API1      & \multicolumn{1}{c|}{0.63} & 0.40 & \multicolumn{1}{c|}{0.62} & \textbf{0.41} & \multicolumn{1}{c|}{0.64} & 0.40 \\ 

API2      & \multicolumn{1}{c|}{0.65} & 0.40 & \multicolumn{1}{c|}{0.64} & 0.40 & \multicolumn{1}{c|}{0.66} & 0.40 \\ 
API3      & \multicolumn{1}{c|}{\textbf{0.66}} & 0.40 & \multicolumn{1}{c|}{0.65} & 0.38 & \multicolumn{1}{c|}{\textbf{0.68}} & 0.40 \\ 
\midrule
Var+API & \multicolumn{1}{c|}{0.65} &\textbf{0.41} & \multicolumn{1}{c|}{\textbf{0.66}} & 0.39 & \multicolumn{1}{c|}{0.66} & 0.40 \\ 
\bottomrule
\end{tabular}
\label{tab: rq1}
\end{table}

\noindent
\textbf{Result:} 
Table~\ref{tab: rq1} shows the result. \tool{} outperforms  the vanilla models for all the settings, all the datasets and all the models.
For Devign data, \tool{} Var1, API3, and VAR+API show 4, 5, and 4 percentage points improvement respectively in terms of the F1-score against the CodeBERT vanilla model. In the UniXcoder models, the improvement for these three approaches are 2, 4 and 2 percentage points respectively. With the GraphCodeBERT model, our approaches API3 and Var+API show 2 and 3 percentage points improvement against the vanilla model. Overall, our approaches show 2-5 percentage points F1 score improvement against the vanilla model. For the Big-Vul dataset, our causal approaches with the CodeBERT model show a 2-3 percentage points improvement, so do the GraphCodeBERT and UniXcoder models.


To the best of our knowledge, these are the best-reported vulnerability detection results in these widely studied datasets~\footnote{our setting keep the inputs in their original form without any perturbation or normalization.}. 
One may suspect that while ignoring spurious features, a model may reduce some of the in-distribution accuracies, as the spurious features also contribute in benign settings. 
In contrast, to our surprise, we find that \tool is learning more significant causal signals, which compensate for the loss of spurious features and also improve in-distribution accuracy.

\begin{question} 
   \textbf{Result:RQ1.}
   \tool outperforms other pre-trained models, suggesting causal learning focuses on the root causes of the vulnerabilities by learning to ignore spurious features. 
   Overall, our causal settings show up to 6\% improvement in F1 in Devign Dataset and 7\% improvement in F1 in Big-Vul dataset.  
\end{question}


\subsection{RQ2: Model Robustness and Generalization}


\noindent
\textbf{Experimental Design.} For robustness evaluation, we compare the performance of the causal models with the vanilla model on the three perturbed datasets presented in Section~\ref{sec:spurious}.
{\it Var1} and {\it Var2} run on the \var dataset, which has the worst performance on the corresponding vanilla model as per Table \ref{tab:var}. 
For example, \var dataset perturbed with Top 5 and Top 10 most frequent variable names perform the worst in vanilla CodeBERT and GraphCodeBERT models, respectively. Hence, we select the Top 5 for comparison with CodeBERT and the Top 10 to compare with GraphCodeBERT. Similarly, {\it API1}--{\it API3} runs on the worst \dead dataset, and {\it Var+API} runs on the worst \joint dataset for the corresponding models.

To investigate the generalization performance of the models, we evaluated the model trained on the Devign dataset using the Big-Vul test dataset (excluding overlapped project FFMPEG). Similarly, we trained on Big-Vul and tested on the Devign dataset. For both experiments, we experimented all the settings in Table~\ref{tab:alg} and used the F1 as metrics.


\noindent
\textbf{Results for Robustness:}  Table \ref{tab:robustness_table} shows the robustness performance in three blocks: the upper block presents the results from running with the worst {\it PerturbVar} data, and similarly, the middle and lower blocks present the results from running with the worst {\it PerturbAPI} and with the worst {\it PerturbJoint} data respectively.

    Between Var1 and Var2, Var1 performs better on Devign data and shows  6, 3, and 2 percentage points improvement in F1 with CodeBERT, GraphCodeBERT and UniXcoder model respectively. For the Big-Vul data, Var1 works better on the Codebert model with 1 percentage point improvement while Var2 is better for the other two models with 4 and 3 percentage points improvement respectively. Overall, both of Var1 and Var2 approaches work better than the vanilla model in terms of robustness. Among API1, API2 and API3, API3 works better in Devign data and demonstrates  10, 5, and 22 percentage points improvement with the three models respectively. For the Big-Vul data, API2 works better with the CodeBERT model and improves by 3 percentage points over CodeBERT vanilla. In GraphCodeBERT, both AP1 and API2 show similar performance and show 2 percentage point improvement. The VAR+API setting shows a similar improvement trend over vanilla performance.

\begin{figure}[htbp]
\centering
\includegraphics[width=0.48\textwidth]{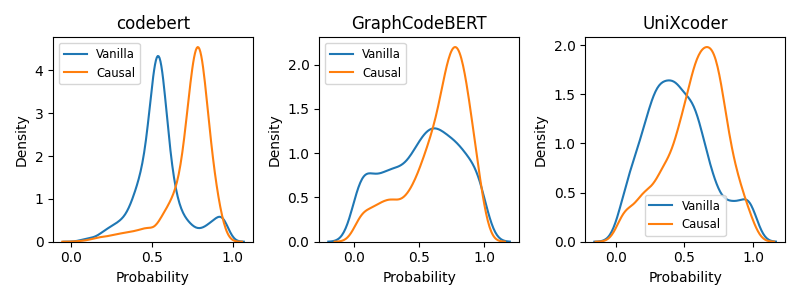}
\caption{Predicted probability density of the Devign  data from Vanilla and Causal approach.}
\label{fig:probability_density_vul}
\end{figure}

In Figure \ref{fig:probability_density_vul}, we show the predicted probability density for the Devign vulnerable data. In each subplot, X-axis is the predicted probability of being vulnerable, and  Y-axis is the count of the examples whose predictions are that probability. The orange lines plot the causal model and the blue lines plot the vanila model. The figure demonstrates that the overall prediction probability for vulnerable data increases, which means the model is more confident in predicting vulnerabilities. Experiment result show that the overall difference of the probability density between vanilla and the causal approach is statistically significant with $p-value<<<0.05.$ with varying effect size, as documented at Table~\ref{tab:cohen}.

We also investigate how many examples from robustness data are predicted incorrectly in Vanilla models like~ Figure 1 and predicted correctly in \tool{}. Table~\ref{tab:counterfactual} shows that \tool{} correctly predict a significant amount of data which are predicted incorrectly in Vanilla models.

\begin{table}[]
\caption{The Robustness performance of \tool{} and the vanilla models.}
\setlength{\tabcolsep}{2pt}
\begin{tabular}{l|cc|cc|cc}
\toprule
Settings  & \multicolumn{2}{c|}{CodeBERT}                          & \multicolumn{2}{c|}{GraphCodeBERT}                     & \multicolumn{2}{c}{UniXcoder}                         \\ \midrule
          & \multicolumn{1}{c|}{Devign} & \multicolumn{1}{c|}{Big-Vul} & \multicolumn{1}{c|}{Devign} & \multicolumn{1}{c|}{Big-Vul} & \multicolumn{1}{c|}{Devign} & \multicolumn{1}{c}{Big-Vul} \\ \midrule
Vanilla   & \multicolumn{1}{c|}{0.52}   & 0.22                     & \multicolumn{1}{c|}{0.58}   & 0.26                     & \multicolumn{1}{c|}{0.52}   & 0.19                     \\\hline
Var1      & \multicolumn{1}{c|}{\textbf{0.58}}   & \textbf{0.23}                     & \multicolumn{1}{c|}{\textbf{0.61}}   & 0.28                     & \multicolumn{1}{c|}{\textbf{0.54}}   & 0.21                     \\ 
Var2      & \multicolumn{1}{c|}{0.55}   & 0.22                     & \multicolumn{1}{c|}{0.58}   & {\bf 0.30}                     & \multicolumn{1}{c|}{0.53}   & \textbf{0.22}                     \\\hline
\midrule
Vanilla   & \multicolumn{1}{c|}{0.52}   & 0.10                     & \multicolumn{1}{c|}{0.47}   & 0.09                     & \multicolumn{1}{c|}{0.35}   & 0.09                     \\\hline
API1      & \multicolumn{1}{c|}{0.52}   & 0.13                     & \multicolumn{1}{c|}{0.35}   & \textbf{0.11}                     & \multicolumn{1}{c|}{0.45}   & \textbf{0.14}                     \\ 
API2      & \multicolumn{1}{c|}{0.56}   & \textbf{0.13}                     & \multicolumn{1}{c|}{0.39}   & \textbf{0.11}                     & \multicolumn{1}{c|}{0.49}   & 0.13                    
                    \\\hline
API3      & \multicolumn{1}{c|}{\textbf{0.62}}   & 0.10                     & \multicolumn{1}{c|}{\textbf{0.52}}   & 0.09                     & \multicolumn{1}{c|}{\textbf{0.57}}   & 0.13  \\\hline
 \midrule
Vanilla   & \multicolumn{1}{c|}{0.52}   & 0.06                     & \multicolumn{1}{c|}{0.45}   & 0.07                     & \multicolumn{1}{c|}{0.22}   & 0.05                     \\\hline
Var+API & \multicolumn{1}{c|}{\textbf{0.55}}   & \textbf{0.07}                     & \multicolumn{1}{c|}{\textbf{0.54}}   & \textbf{0.08}                     & \multicolumn{1}{c|}{\textbf{0.31}}   & \textbf{0.10}                     \\ 
\bottomrule
\end{tabular}
\label{tab:robustness_table}
\end{table}

\begin{table}[]

\caption{Cohen's d effect size of the difference of the probability density between Vanilla and Causal approach for vulnerable data.}
\setlength{\tabcolsep}{2pt}
\begin{tabular}{l|cc|cc|cc}
\toprule
Setting & \multicolumn{2}{c|}{CodeBERT}                           & \multicolumn{2}{c|}{GraphCodeBERT}                      & \multicolumn{2}{c}{UniXcoder}                          \\ \toprule
        & \multicolumn{1}{l|}{Devign}  & \multicolumn{1}{l|}{Big-Vul} & \multicolumn{1}{l|}{Devign}  & \multicolumn{1}{l|}{Big-Vul} & \multicolumn{1}{l|}{Devign}  & \multicolumn{1}{l}{Big-Vul} \\
\midrule
var     & \multicolumn{1}{c|}{trivial} & midium                   & \multicolumn{1}{c|}{trivial} & small                    & \multicolumn{1}{c|}{trivial} & small                    \\ 
api     & \multicolumn{1}{c|}{large}   & midium                   & \multicolumn{1}{c|}{small}   & small                    & \multicolumn{1}{c|}{large}   & small                    \\ 
combine & \multicolumn{1}{c|}{large}   & small                    & \multicolumn{1}{c|}{large}   & small                    & \multicolumn{1}{c|}{small}   & midium                   \\ 
\bottomrule
\end{tabular}
\label{tab:cohen}
\end{table}

\begin{table}[]
\caption{Number of pertubed examples whose predictions are incorrect in Vanilla model (see Figure~ 1) but correct in \tool.}
\begin{tabular}{l|c|c|c}
\toprule
Dataset & CodeBERT & GraphCodeBERT & UniXcoder  \\ \midrule
Devign  & 298      & 255           & 368         \\ \midrule 
Big-Vul     & 205      & 93            & 130        \\ 
\bottomrule
\end{tabular}
\label{tab:counterfactual}
\end{table}

\begin{table}[]
\setlength{\tabcolsep}{2pt}
\caption{The generalization performance of \tool{} and the vanilla models.}.
\label{tab:generalization}
\begin{tabular}{l|cc|cc|cc}
\toprule
Settings                          & \multicolumn{2}{c|}{CodeBERT}      & \multicolumn{2}{c|}{GraphCodeBERT} & \multicolumn{2}{c}{UniXcoder}     \\\midrule
\cellcolor[HTML]{FFFFFF}          & \multicolumn{1}{c|}{Devign} & Big-Vul  & \multicolumn{1}{c|}{Devign} & Big-Vul  & \multicolumn{1}{c|}{Devign} & Big-Vul  \\ \midrule
\cellcolor[HTML]{FFFFFF}Vanilla   & \multicolumn{1}{c|}{0.11}   & 0.08 & \multicolumn{1}{c|}{0.11}   & 0.10 & \multicolumn{1}{c|}{0.10}   & 0.07 \\ \midrule
\cellcolor[HTML]{FFFFFF}Var1      & \multicolumn{1}{c|}{0.12}   & 0.17 & \multicolumn{1}{c|}{0.11}   & 0.11 & \multicolumn{1}{c|}{\textbf{0.12}}   & 0.13 \\ 
\cellcolor[HTML]{FFFFFF}Var2      & \multicolumn{1}{c|}{0.12}   & \textbf{0.18} & \multicolumn{1}{c|}{0.11}   & \textbf{0.14} & \multicolumn{1}{c|}{\textbf{0.12}}   & 0.15 \\ 
\midrule
\cellcolor[HTML]{FFFFFF}API1      & \multicolumn{1}{c|}{0.12}   & 0.17 & \multicolumn{1}{c|}{0.11}   & 0.12 & \multicolumn{1}{c|}{\textbf{0.12}}   & 0.11 \\ 

\cellcolor[HTML]{FFFFFF}API2      & \multicolumn{1}{c|}{0.12}   & 0.16 & \multicolumn{1}{c|}{0.11}   & \textbf{0.14} & \multicolumn{1}{c|}{\textbf{0.12}}   & 0.18 \\ 
\cellcolor[HTML]{FFFFFF}API3      & \multicolumn{1}{c|}{\textbf{0.13}}   & 0.15 & \multicolumn{1}{c|}{0.11}   & 0.10 & \multicolumn{1}{c|}{\textbf{0.12}}   & \textbf{0.21} \\
 \midrule
\cellcolor[HTML]{FFFFFF}Var+API & \multicolumn{1}{c|}{0.12}   & 0.17 & \multicolumn{1}{c|}{0.11}   & 0.12 & \multicolumn{1}{c|}{\textbf{0.12}}   & 0.18 \\ 
\bottomrule
\end{tabular}
\end{table}

\textbf{Results for Generalization:} We present the generalization results in Table \ref{tab:generalization}. The Devign column defines the F1 score of the Big-Vul test set when evaluated on the model trained with the Devign train set. Similarly, the Big-Vul column presents the F1 score of the Devign test set evaluated on the model that is trained with the Big-Vul train set. Our results show that when the model is trained on the Devign train set and the Big-Vul test set is used as the out-of-distribuition (OOD), our causal approach shows 1-2 percentage points improvement for CodeBERT and UniXcoder models.   But when the Devign test set is used as the OOD data on the model trained with Big-Vul data, CodeBERT model shows 7-10 percentage points improvement, GraphCodeBERT model shows at most 4 percentage points improvement and the UniXcoder model shows 6-14 percentage points improvement.

In this RQ, we show that causal learning has the potential of significantly improving the robust accuracy and generalization. 
In such a setting, since the spurious features may not be present in the evaluation data, learning to ignoring them helps \tool to significantly improve the performance.

\begin{question} 
   \textbf{Result:RQ2.} \tool{} shows up to 62\% and 100\% improvement in Devign and Big-Vul robustness data respectively. \tool{} also improves the generalization performance up to 100\% and 200\% for both datasets respectively.
\end{question}

\subsection{RQ3: Ablation Studies}

\noindent
\textbf{Experiment Design.} In this RQ, we investigate the design choices for \tool{}. In the first setting, we use K=1 and set $x'=x$ in Algorithm 2. Here, we investigate if we don't use marginalization, how our approach performs. We evaluated the models on the robustness testing data. Due to space, we show the results for the setting of {\it Var+API}.

In the second setting, we investigated how our approach performs when using different early layers to represent $M$. Hence, we extract M from the first, second, third, and fourth layers and use that M in our causal approach respectively.
\mycomment{
\begin{itemize}[leftmargin=*]
    \item {\it K=1:} We investigate if we don't use marginalization, then how our approach performs. For this purpose, we select K=1 and set $x'=x$ to evaluate the models on the data {that are used for robustness testing}.
    \item {\it Encoding M:} In this experiment, we investigate how our approach performs when using different early layers to represent $M$. Hence, we extract M from the first, second, third, and fourth layers and use that M in our causal approach respectively.
\end{itemize}
}

\begin{figure}[htbp]
\centering
\includegraphics[width=0.48\textwidth]{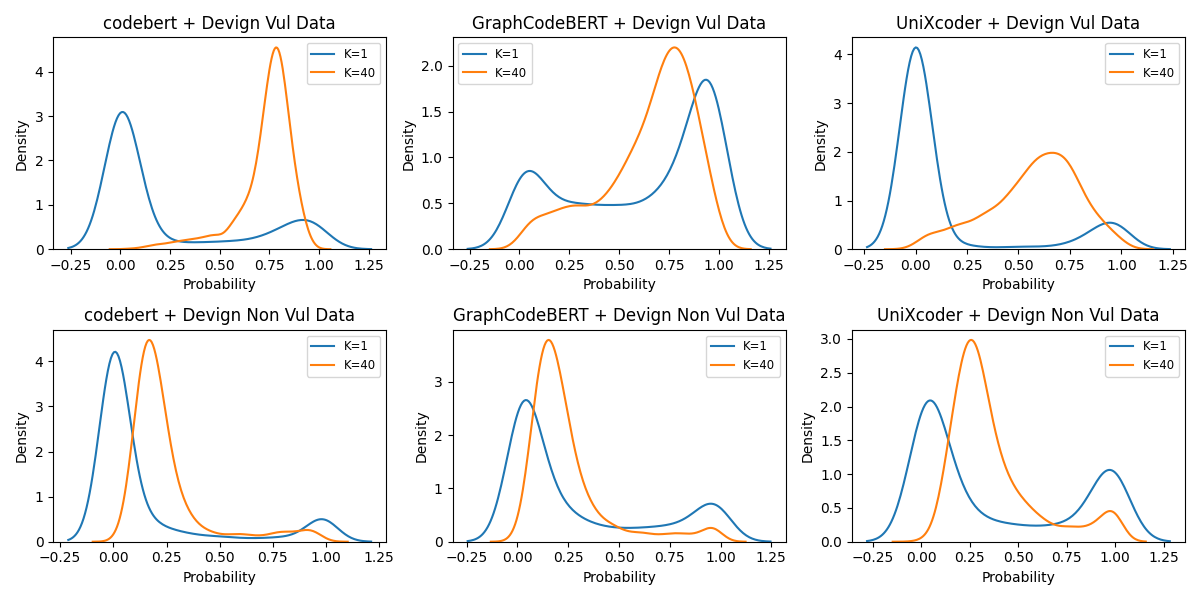}
\caption{Prediction Probability Density of \tool for K=1 and K=40. }
\label{fig:prob_k_1vs40}
\end{figure}

\noindent
\textbf{Results.} In Figure \ref{fig:prob_k_1vs40}, we used the probability density plots similar to Figure~\ref{fig:probability_density_vul}. The orange lines plot K=40, and the blue lines plot K=1. From the results, we can clearly see that for both vulnerable and non-vulnerable labels, K=40 learned better and reported more confident predictions towards ground truth labels.

Table \ref{tab:ablationel} demonstrates the result of using different early layers to extract $M$. We choose the {\it Var+API} settings for all models to present the result. For all the models and datasets, layer four reported the best performance. 

\begin{table}[]

\caption{The performance of the Causal Approach when different early layer is used.}
\begin{tabular}{l|c|c|c}
\toprule
Model         & Early Layer          & \begin{tabular}[c]{@{}c@{}}Devign\\ (Var+API)\end{tabular} & \begin{tabular}[c]{@{}c@{}}Big-Vul\\(Var+API)\end{tabular} \\
\midrule
CodeBERT      & 1  & 0.6494   & 0.4034   \\
CodeBERT      & 2 & 0.6528  & 0.4018    \\
CodeBERT      & 3                    & 0.6501                                                       & 0.4026                                                    \\
CodeBERT      & 4                    & \textbf{0.6546}                                                       & \textbf{0.4055}                                                    \\
\midrule
GraphCodeBERT & 1                    & 0.6156                                                       & 0.3916                                                    \\
GraphCodeBERT & 2                    & 0.6170                                                       & 0.3903                                                    \\
GraphCodeBERT & 3                    & 0.6363                                                       & 0.3908                                                    \\
GraphCodeBERT & 4                    & \textbf{0.6570}                                                       & \textbf{0.3927}                                                    \\
\midrule                            
UniXcoder     & 1                    & 0.5728                                                       & 0.4049                                                    \\
UniXcoder     & 2                    & 0.5720                                                       & 0.4055                                                    \\
UniXcoder     & 3                    & 0.5550                                                       & 0.4056                                                    \\
UniXcoder     & 4                    & \textbf{0.6609}                                                       & \textbf{0.4058}  \\
\bottomrule
\end{tabular}
\label{tab:ablationel}
\end{table}


\begin{question} 
   \textbf{Result:RQ3.} Our results show that marginalization (backdoor criterion) helps the model to focus on the causal features instead of spurious features. On the other hand, we found Layer four is better to use to compute $M$ from $x'$ than the early three layers.
\end{question}

%% file: body/7.threats.tex
\section{Threats to Validity}
\label{sec:threats}
To discover spurious features, our experiment design follows the literature ~\cite{data_avail} and ensures our perturbation follows consistency, naturalness, and semantic-preservation. 

Deep learning models may report improved results due to a better random seed. All of our experiments have been run with three random seeds. Our causal models have consistently shown improvement across all the datasets and all the models. We have done a statistical test to show that our improvement is statistically significant. In addition to F1, our probability density plots shown in Figures~\ref{fig:probability_density_vul} and ~\ref{fig:prob_k_1vs40} also strongly demonstrated our improvement. 

The causal learning makes the assumption that the code representation $R$ learned the causal features. Although we are not sure if that's the case, we see the improvement of our results in all settings.

Our evaluation worked on two real-world vulnerability datasets, including both balanced and imbalanced data, and the three SOTA models. In the future, we plan to experiment with more datasets and models.

\mycomment{
\noindent\textbf{Discovering spurious features:} While we aim to perturb our input in a way that respects consistency, naturalness, and semantic-preservation, there are some limitations at hand. A small subset of the code samples are outliers in size (consider one-line functions).
For these one-liners, injecting dead-code in multiple locations may threaten naturalness and plausibility, as it may serve to overwhelm the sample. Let us emphasize, however, that 1) ultimately, one-liners represent a negligible percentage of the test set’s distribution, and 2) our main focus is to demonstrate the causal model’s ability to undo the spurious features, irrespective of the perturbed samples' plausibility. 

\noindent\textbf{Causal learning experiments}
Our representation learning in implementation may fail to capture the causes: assumption, just ask models to not use these spurious features

In algorithm 2, we select K=40, based on the initial experiment where we evaluated the performance on K=20, 40 and 60. As the evaluation using different K is a costly process, we select K=40 based on these three experiments. There can be other K that could show more improvement. However, K=40 already shows better improvement which proves our hypothesis. 

We didn't try with other layers 5-12 to extract M from $x'$. However, layer four already shows improvement. Also, as the initial experiment says the performance of layer six is almost the same to layer four, we believe the other layers will also be consistent like layer six.

In the LineVul \cite{LineVul} paper, the performance of Big-Vul data is 91\% using the LineVul model. However, if we use the same preprocessing technique used by the CodeBERT, UniXcoder, and GRaphCodeBERT, the performance of the LineVul model on Big-Vul data drops to 36-38\%.
}

%% file: body/6.related.tex
\section{Related Work}
\label{sec:related}

\noindent{\bf Deep learning for Vulnerability}:  Deep learning vulnerability models can be separated into graph neural-network (GNN) based or transformer-based models. GNN-based models capture AST, control-flow, and data-flow information into a graph representation. Recent GNN-based models ~\cite{linevd, mvd, deepvd, zhou2019devign, chakraborty2020deep}, have proposed statement-level vulnerability prediction. 

In contrast, the transformer models are pre-trained in a self-supervised learning setting. They can be categorized by three different designs: encoder-based, decoder-based ~\cite{chen2021evaluating}, and hybrid architectures ~\cite{ahmad2021unified, guo2022unixcoder, data_avail} that combine elements from both approaches. Encoder-based models such as CodeBERT ~\cite{guo2020graphcodebert, feng2020codebert} often employ the masked-language-model (MLM) pre-taining objective; some are coupled with a contrastive learning approach ~\cite{bui2021corder, ding2022towards}, while others aim to make pre-training more execution aware ~\cite{ding2023traced}, bi-modal ~\cite{ahmad2021unified} or naturalized ~\cite{data_avail}. Vulnerability detection has been one of the important downstream tasks for these models. In this paper, we used three recent SOTA transformer-based models: GraphCodeBERT, CodeBERT, and UniXcoder and show that causality can further improve their performance. 

There have been also studies for vulnerability detection models regarding their robustness and generalization. In recent work, Steenhoek et al. ~\cite{steenhoek2023empirical} evaluated several SOTA vulnerability models to assess their capabilities to unseen data. Furthermore, they touch on spurious features and found some tokens such as "error" and "printf" are frequently used to make predictions, which can lead to mispredictions.
In another systematic investigation of deep learning-based vulnerability detection, Chakraborty et al.~\cite{chakraborty2020deep} stated that vulnerability detection models did not pick up on relevant code features, but instead rely on irrelevant aspects from the training distribution (such as specific variables) to make predictions. Our work designed novel perturbations to confirm the hypothesized spurious features, and we found different names are used for different labels as spurious features. None of the existing work have conducted such studies.

\noindent{\bf Causal learning in SE}: To our knowledge, applying causality is relatively new in SE. Cito et al.~\cite{2022icsecito} is the most relevant recent work that investigates perturbations on source code which cause a model to “change its mind”. This approach uses a masked language model (MLM) to generate a search space of "natural" perturbations that will flip the model prediction. These natural perturbations are called “counterfactual explanations.” Our work also seeks for natural perturbations, and uses variable names and API names in programs to perform the perturbation. However, our work is different in that we require our perturbation to be semantic-preserving. Furthermore, our goal of promoting models to flip their decision is to discover spurious features, instead of explaining the cause of a bug.
There has also been orthogonal work that uses counterfactual causal analysis in the context of ML system debugging ~\cite{unicorn2022, zhong2022fusion}. Unlike our work, these approaches do not use the backdoor criterion to remove spurious features. 
 
\noindent{\bf Causal learning in Other Domains}: Our work drew inspirations from Mao et al.~\cite{2022cvprmao}. This work addresses robustness and generalization in the vision domain using causal learning. They use "water bird" and "land bird" as two domains and show that by applying the backdoor criterion, the causal models can learn "invariants" of the birds in two domains and achieve better generalization. Their work does not discover spurious features, and their causal learning does not target spurious features. 



%% file: body/8.conclusion.tex
\section{Conclusions and Future Work}
\label{sec:conclusion}
This paper proposed the first step towards causal vulnerability detection. We addressed several important challenges for deep learning vulnerability detection. First, we designed novel perturbations to expose the spurious features the deep learning models have used for prediction. Second, we formulate the problem of deep learning vulnerability detection using {\it causality} and {\it do calculus} so that we can apply causal learning algorithms to remove spurious features and push the models to use more robust features. We designed comprehensive experiments and demonstrated that \tool{} improved accuracy, robustness and generalisation of vulnerability detection. In the future, we will plan to discover more spurious features and explore the causal learning for other applications in software engineering.

\section{Acknowledgements}

We thank the anonymous reviewers for their valuable feedback. This research is partially supported by the U.S. National Science Foundation (NSF) under Award ~\#2313054.